\documentclass[twocolumn,epjc3]{svjour3}          
\usepackage{graphicx}
\usepackage{slashed}
\usepackage{amssymb,amsopn}
\usepackage[cmtip,arrow]{xy}
\usepackage{pb-diagram,pb-xy}
\usepackage{slashed}
\def \be {\begin{equation}}
\def \ee {\end{equation}}
\def \bea {\begin{eqnarray}}
\def \eea {\end{eqnarray}}

\def \dels {\partial\kern-.5em / \kern.5em}
\def \As {{A\kern-.5em / \kern.5em}}
\def \Ds {D\kern-.7em / \kern.5em}

\newcommand{\sla}[1]{\hbox{{$#1$}\llap{$/$}}}
\def \be {\begin{equation}}
\def \ee {\end{equation}}
\def \bea {\begin{eqnarray}}
\def \eea {\end{eqnarray}}
\RequirePackage[T1]{fontenc}
\smartqed  
\RequirePackage{graphicx}
\RequirePackage{mathptmx}      
\RequirePackage{flushend}
\RequirePackage[numbers,sort&compress]{natbib}
\RequirePackage[colorlinks,citecolor=blue,urlcolor=blue,linkcolor=blue]{hyperref}
\journalname{Eur. Phys. J. C}
\begin{document}
\title{Microcausality and quantization of the fermionic Myers-Pospelov model}
\subtitle{}
\author{Justo Lopez-Sarrion\thanksref{e1,addr1}
\and Carlos M. Reyes\thanksref{e2,addr2} }

\thankstext[$\star$]{t1}{}
\thankstext{e1}{e-mail: justinux75@gmail.com}
\thankstext{e2}{e-mail: creyes@ubiobio.cl}

\institute{Departamento de F{\'i}sica,
Universidad de Santiago de Chile, Casilla 307,
Santiago, Chile.\label{addr1}
          \and
          Departamento de Ciencias
B{\'a}sicas, Universidad del B{\'i}o-B{\'i}o, Casilla 447,
Chill\'an, Chile.\label{addr2}}
\date{Received: date / Accepted: date}
\maketitle
\begin{abstract}
We study the fermionic sector of the Myers and Pospelov theory with
a general background $n$. The spacelike case without temporal
component is well defined and no new ingredients came about, apart
from the explicit Lorentz invariance violation. The lightlike case is
ill defined and physically discarded. However, the other case where
a nonvanishing temporal component of the background is present, the
theory is physically consistent. We show that new modes appear as a
consequence of higher time derivatives. We quantize the timelike
theory and calculate the microcausality violation which turns out to
occur near the light cone.
\end{abstract}
\section{Introduction}
The need for a more fundamental theory at high energies has been
justified in many different contexts. Divergences in quantum field theory,
singularities in gravity and the lack of a unified quantum framework
for all forces, are some of them. A consequence arising from this
consideration, which has been extensively studied, is the
possibility of having Lorentz invariance violation in the form
of effective corrections \cite{KOSTELECKY1}. This idea naturally
leads to new extensions of the standard model and modified
dispersion relations for particles. Today experimental searches for
Lorentz invariance violation are being carried in diverse frontiers
\cite{RL1}.

In this context the Myers-Pospelov theory is a model that
introduces Lorentz invariance violation through dimension five
operators \cite{MP,BM}. The breakdown of Lorentz symmetry takes
place in the scalar, fermion and gauge sectors and is characterized
by an external timelike four-vector $n_{\mu}$ defining a preferred
reference frame. Experimental bounds for this model have been
studied in several phenomena, such as synchrotron radiation
\cite{syn}, gamma ray bursts \cite{grb}, neutrino physics
\cite{NEUT}, radiative corrections \cite{11,12}, generic backgrounds
\cite{21,MR}, and others \cite{others}. Typically, these
phenomenological studies assume $n$ to lie purely in the temporal
direction \cite{berto}. In this work we will take $n$ as general as possible and
eventually we will consider some special choices.

In recent years, theories with higher time derivatives have been
proposed as extensions of the standard model of particles \cite{BG}.
One of the main advantages is that these theories soften the
ultraviolet behavior of the quantum field theory, and hence problems
like the hierarchy puzzle seem to be solved. Although they contain
negative norm states \cite{GHOSTS,P-U} the theoretical consistency
was established many years ago \cite{INDFMETRIC}. It can be shown
that although unitarity is maintained, the price to pay is the lost
of causality \cite{CUTW}.

The new negative norm modes are relevant at high energies screening
the ultraviolet effects of any standard quantum field theory 
leading to a low energy limit
which is not sensitive to the details of the effective theory at
microscopic scales (see, however
\cite{examples}). The Myers and Pospelov theory has these
ingredients when $n$ has a nonvanishing temporal component. Hence,
it is interesting to investigate the role of these new modes in
order to check the behavior of the low energy limit of
Myers-Pospelov theory. In this work we will analyze how these new
modes affect the quantization of the theory, because it is the first
step to study such low energy limit.

Moreover, interacting theories with higher time derivatives lose
causality at the microscopic level if we want to maintain unitarity.
An effect of this acausal behavior is for instance the negativity
of certain decay rates. But also the Lorentz violating Myers and
Pospelov theories have a natural violation of the microcausality
principle, even without interactions \cite{klink}. Since in this work we will not
deal with interactions we will focus on the study of the last source
of violation of microcausality.

The layout of this work is the following. In Sect. \ref{Sec2} we introduce
the fermionic Myers and Pospelov model where we find the dispersion
relation in an arbitrary background. For special choices of the
preferred four-vector we analyze the causality and stability of the
different theories. In Sect. \ref{Sec3} we review the main aspects of a
higher time derivative theory like the fermionic Lee-Wick model
which will help us to understand the remaining sections. In Sect. \ref{Sec4}
we quantize the timelike Myers and Pospelov theory by performing a
decomposition of the theory into four individual fermionic
oscillators. In Sect. \ref{Microcausality} we discuss violations of microcausality where
a perturbative computation of the anticommutator function is given.
In the last section we give the conclusions and final comments. In \ref{app} 
we characterize the general solutions
and dispersion relations.
\section{Fermionic Myers-Pospelov model}\label{Sec2}
The fermionic sector of the Myers-Pospelov theory is
given by the Lagrangian
\begin{eqnarray}\label{M-P-Lag1}
\mathcal L=\bar \psi(i {\sla{\partial}}-m)\, \psi+ \bar
\psi \, \sla{n}(g_1+g_2\gamma_5) \, (n \cdot \partial)^2 \psi,
\end{eqnarray}
where $g_1$ and $g_2$ are inverse Planck mass dimension couplings
constants and $n$ is a dimensionless
four-vector defining a preferred reference frame with $n^2=+1,-1,$ or $0$.

The variation of the Lagrangian (\ref{M-P-Lag1}) produces the
equations of motion
\begin{eqnarray} \label{eqm}
 \left[i\slashed\partial -m+g_1\slashed n
 (n\cdot\partial)^2 +g_2 \slashed n\gamma^5
 (n\cdot\partial)^2\right]\psi(x)=0.
\end{eqnarray}
In momentum space, $\psi (x)=\int d^4p \,e^{-i p\cdot x} \,\psi(p)$,
we obtain an algebraic equation,
\begin{eqnarray} \label{eqm.momentum}
 \left[\slashed p -m-g_1\slashed n
 (n\cdot p)^2 -g_2 \slashed n\gamma^5
 (n\cdot p)^2\right]\psi(p)=0.
\end{eqnarray}
The dispersion relation is given by
\begin{eqnarray} \label{reldisp-cuad2}
\Big(p^2-m^2-2g_1 (n \cdot p)^3
 + n^2(g_1^2-g_2^2)(n \cdot p)^4\Big)^2\nonumber \\
 -4 (n \cdot p)^4 g_2^2\Big((n \cdot p)^2-p^2n^2\Big)=0.
\end{eqnarray}
In general (\ref{reldisp-cuad2}) is is an eighth order polynomial in
$\omega$ and it would yield at most eight real solutions. However,
if $n_0= 0$ the order of the polynomial in $\omega $ is four
corresponding to particles and antiparticles of spin $1/2$. The
negative solutions correspond to antiparticles modes while the
positive ones are particles modes. The situation
 for $n_0\neq 0$ is to obtain twice the number
of solutions than in the standard case. This is due to the fact that
we are dealing with a theory with
 higher time derivatives as can be
seen from the equation of motion (\ref{eqm}).
In the next subsection we will discuss
in more detail the nature of these extra solutions.

A derivation of Eq. (\ref{reldisp-cuad2}) and the eigenspinor
solutions are given in the Appendix. In what follows we will
consider the case $g_2=0$. The case of a nonvanishing $g_2$
introduces very complicated parameterizations as it can be seen in
the Appendix. However, it does not contribute to new relevant
features and renders the calculations cumbersome. The reader
interested in this case can go through the Appendix.
\subsection{The timelike model}\label{2a}
We start to analyze the purely timelike case by taking $n=(1,0,0,0)$
and as mentioned above setting $g_2=0$. In this case the dispersion
relation (\ref{reldisp-cuad2}) reduces to
\begin{eqnarray} \label{disp.rel-timelike}
\omega^2-\vec {p}^2-m^2-2g_1\omega^3+g_1^2 \omega^4=0,
\end{eqnarray}
from where we obtain the four solutions
\begin{eqnarray}\label{field-energy2}
\omega_{(a=1,2)}=  \frac{1- \sqrt{1-4(-1)^{a}g_1E_{\vec p}}}{2g_1},\nonumber \\
\omega_{(a=3,4)}=  \frac{1+ \sqrt{1+4(-1)^{a} g_1E_{\vec p}}}{2g_1},
\end{eqnarray}
with $E_{\vec p}=\sqrt{\vec p^2+m^2}$.

The solutions $\omega_{1,2}$ in the limit $g_1 \to 0$ tend to the
usual solutions $\mp E$ while the solutions $\omega_{3,4}$ go to
infinity. These singular solutions are called Lee-Wick modes
\cite{INDFMETRIC} and will be explained in more detail in the next
section.

In order to see the qualitative behavior of the solutions let us
define the two functions $f(\omega)=
\omega^2-m^2-2g_1\omega^3+g_1^2 \omega^4 $ and $g(\omega)=\vec p^2$,
and plot these functions of $\omega$ in Fig.\ 1.
\begin{figure}
\centering
\includegraphics[width=0.47\textwidth]{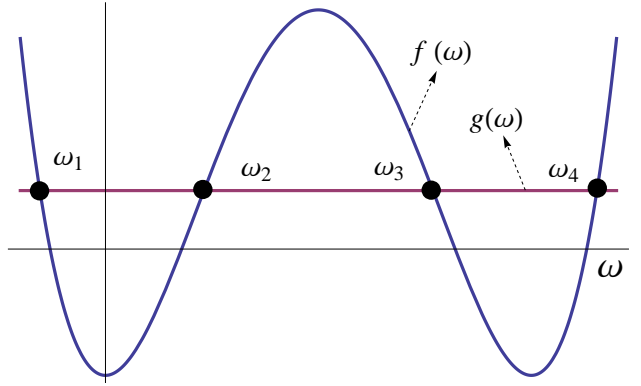}
\caption{The intersection of the horizontal straight line $g(\omega)=\vec p^2$ with
the curve $f(\omega)$
corresponds to the solutions $\omega_a$ given in 
 (\ref{field-energy2}).}
\end{figure}\label{fig1}
The solutions are the intersection points of the curve $f$ and
the horizontal straight line corresponding to the fixed value of the
momentum square, \emph{i.e.} $g(\omega)=\vec p^2$. Hence, for small
values of $\left|\vec p\right|$ we find four solutions, one negative frequency
which corresponds to an antiparticle and three positive frequencies.
Among the positive frequencies the smallest one is the normal
particle frequency and the other two correspond to Lee-Wick modes.
It is peculiar the behavior of the Lee-Wick solution whose
frequency decreases with momentum, this will continue until the
momentum reaches the value of $  \left|\vec
p\right|_{max}=\sqrt{\frac{1}{16g_1^2}-m^2}$ where it collapses with
the normal particle mode. Above these values the solutions $\omega_2$
and $\omega_3$ become complex introducing stability problems.
Furthermore, it is worth noting the differences in energy between
particles and antiparticles which in the limit $mg_1<<1$ turns out
to be $4 \left|g_1 \right| m^2$.
\begin{figure}
\centering
\includegraphics[width=0.47\textwidth]{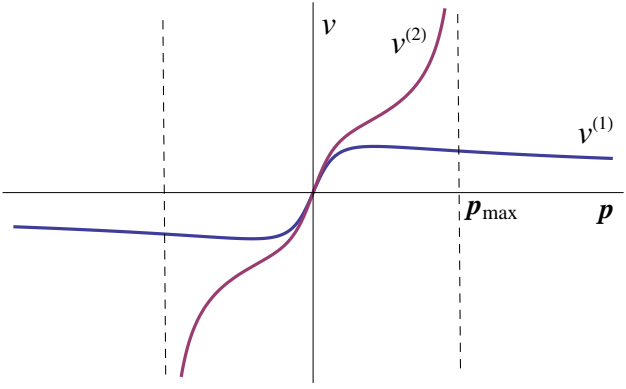}
\caption{The magnitude of the
group velocities $v^{(1)}$ and $v^{(2)}$
given in (\ref{group-velocity}).}
\end{figure}\label{fig2}

Some insight can be gained into the possible violations of
microcausality in the model by looking at the group velocities
\cite{RL}. The magnitude of the group velocities are
\begin{eqnarray} \label{group-velocity}
v_{(a=1,4)}^{(1)}&=&(-1)^a\frac{  \left|   \vec p
\right|}{E_{\vec p}\sqrt {1+4g_1E_{\vec p}}},\nonumber
\\
v_{(a=2,3)}^{(2)}&=&(-1)^a\frac{  \left|
\vec p  \right|}{E_{\vec p}\sqrt {1-4g_1E_{\vec p}}},
\end{eqnarray}
and they are plotted in Fig.\ 2. According to the criteria of
\cite{RL} we should expect small violations of microcausality in the
theory since the velocities $v_{(a=3,4)}^{(2)}$ can exceed normal
signal propagation at high momenta. In section \ref{Microcausality}
we give a detailed computation of microcausality.
\subsection{The lightlike model}
In the lightlike case and for simplicity taking $n_0=1$ the
dispersion relation reads
\begin{eqnarray} \label{disp.rel-lightlike1}
\omega^2-\vec p^2-m^2-2g_1  (\omega-\left|\vec p \right|\cos \theta)^3=0,
\end{eqnarray}
where $\theta$ is the angle between $\vec n$ and $\vec p$.
The solutions are
\begin{eqnarray} \label{disp.rel-lightlike}
\omega_1&=&\frac{1}{6g_1}+ \left| \vec p\right| \cos \theta-A,\nonumber
\\
\omega_2&=&\frac{1}{6g_1}+\left| \vec p\right|\cos \theta+B,\nonumber
\\
\omega_3&=&\frac{1}{6g_1}+ \left| \vec p\right| \cos \theta+B^{*},
\end{eqnarray}
with
\begin{eqnarray}
A=\frac{1+12g_1
 \left| \vec p\right| \cos \theta}{6g_1 K^{1/3}}
+\frac{K^{1/3}}{6g_1},
\end{eqnarray}
\begin{eqnarray}
B&=&\frac{ (1+i\sqrt{3})
(1+12g_1 \left| \vec p\right|\cos \theta)}{12g_1 K^{1/3}} \\
&&
+\frac{(1-i\sqrt{3}) K^{1/3}}{12g_1},\nonumber
\end{eqnarray}
and
\begin{eqnarray}
K&=&-1+54E_{\vec p}^2g_1^2 -18 g_1 \left| \vec p\right|\cos \theta      \\ &&
\times(1+3g_1 \left| \vec
 p\right|\cos \theta)+3\sqrt 3 \left| g_1\right|
\sqrt{-\Delta}\nonumber
\end{eqnarray}
where $\Delta$ is the discriminant of the third order polynomial (\ref{disp.rel-lightlike1}).

Here, the roots can be real or
complex depending on whether the discriminant is greater or less than zero,  
respectively. 
Therefore, the quantization of this model presents an
extra complication of instability due to complex solutions.
To see this more clearly consider the discriminant up to the linear order 
\begin{eqnarray}
\Delta\approx4E_{\vec p}^2  (1+18g_1\left|\vec p \right|\cos \theta).
\end{eqnarray}
For example we see that for momenta higher than $\left|\vec p
\right|_{max}=\frac{1}{18g_1\left|\cos \theta\right|}$ 
the solutions in the anti-parallel direction can be imaginary.
For these very high momenta the theory can violate causality since the
retarded Green function gives a contribution at times $t<0$. 
This is very similar to what occurs in the
timelike model for $\omega_{2,3}$, see
Fig.\ 3 of Sect 5, however here the instabilities are
not controllable by restricting to lower momenta or introducing a
cutoff \cite{11}.
\subsection{The spacelike model}
Without loss of generality we
can take the preferred vector as $\vec n=(0,0,1)$.
The dispersion relation for this case is
\begin{eqnarray}
\omega^2-E_{\vec p}^2+2g_1 p_z^3
-g_1^2p_z^4=0.
\end{eqnarray}
The frequency solutions are
\begin{eqnarray}\label{spacelike-solutions}
\omega_{\pm}= \pm \sqrt{ p_{x}^2+p_{y}^2+(p_z-g_1 p_z^2 )^2+m^2}.
\end{eqnarray}
Note that these solutions are always real and that we recover 
the usual dispersion
relation 
when the preferred vector is orthogonal to the 
propagation, called blind momenta directions.

To discuss the causal structure of the theory let us compute the
retarded Green function. We must check that it vanishes for times before
the interaction is turned on, that is to say, before the time $t=0$. 
The retarded Green function in this
case is
\begin{eqnarray}
iS_R(x)&=&(i\sla{\partial}-g_1\gamma^z \partial_z^2+m)\nonumber \\
&&\times
\int_{C_R}\frac{d^4p}{(2\pi)^4}\frac{e^{-ip\cdot x}}{(p_0^2-\omega^2)},
\end{eqnarray}
where the poles are given by the solutions
(\ref{spacelike-solutions}) and $C_R$ is the contour above the real
axis as depicted in Fig\ 3 of Sect 5. The argument that causality is preserved
is rather simple and goes as follows. For times
$t<0$ the contour $C_R$ must be closed from above and 
therefore does not enclose any pole. Recall
that the poles lie on
the real axis even for arbitrary high momenta. In this way 
 there are no violations of causality in
the spacelike model.
\section{Lee-Wick theories}\label{Sec3}
Before facing the problem of quantization, let us review some
general aspects concerning higher derivative theories which may not
be familiar for some readers. These kind of theories were studied by
Lee and Wick and others some decades ago \cite{P-U,INDFMETRIC,CUTW}
and recently there has been a growing interest in them regarding the
hierarchy problem in the standard model \cite{BG}. Unlike the theory
we are considering, the Lee-Wick models are Lorentz invariant
theories, however, they have in common the higher order time
derivatives. We will devote this section to summarize the main
features of the fermionic sector of a Lee-Wick model which will be
important for our subsequent analysis.

In particular let us consider the Lagrangian
\begin{eqnarray}\label{M-P-Lag2}
\mathcal L=\bar \psi(i {\slashed{\partial}}-m)\,
\psi-\frac{g}{\Lambda}  \bar \psi \,\Box\, \psi,
\end{eqnarray}
where $g$ is a dimensionless positive coupling constant and
$\Lambda$ is an ultraviolet energy scale.

By defining the new fields
\begin{eqnarray}
 \psi_{+} & =&\beta (i
 {\sla{\partial}}+m_{-})\, \psi,\nonumber \\
 \psi_{-}  &=& \beta
 (i {\sla{\partial}}-m_{+})\, \psi,
\end{eqnarray}
with $\beta= \left(\frac{g/\Lambda}{ m_{+}+m_{-}}\right)^{\frac{1}
{2}}$ and
\begin{eqnarray}
m_\pm = \frac{\mp 1 +\sqrt{1+4g\frac{m}{\Lambda}}}{2g}\Lambda,
\end{eqnarray}
the Lagrangian (\ref{M-P-Lag2}) can be written in terms of these
fields as
\begin{eqnarray}
\mathcal L&=&\bar \psi_{+}  (i {\sla{\partial}}-m_{+})\,
\psi_{+}-\bar \psi_{-}  (i {\sla{\partial}}+m_{-})\, \psi_{-}.
\end{eqnarray}
Here we have written a higher time derivative theory in terms of to
two decoupled standard fermions. However, the second mode has the
wrong sign in fronts of its Lagrangian density.

The non vanishing anticommutators will be
\begin{eqnarray}
\{\psi^{\alpha}_+(\vec x,t),\psi^{\dag \beta}_+(\vec y,t)\}&=&-\{\psi_-^{\alpha}(\vec
x,t),\psi^{\dag \beta}_-(\vec y,t)\}\nonumber \\
&=&\delta^{\alpha \beta}\delta^3(\vec x-\vec y).
\end{eqnarray}
Note that the minus sign of the anticommutators of the minus fields
is responsible for the negative norm states. 

Now, decomposing the new fields in terms of plane wave solutions we find 
\begin{eqnarray}
  \psi_ {+}  (\vec x,t)&=&\sum_{s}\int \frac{d^3p }{(2\pi)^3}\frac{1}
  {\sqrt{2 E_{+}}} \\
  &&\times \left[b_{+}^s({\bf p})e^{-ip_{+}\cdot x}u_{+}^s({\bf p})+
  d^{s\dag}_{+}({\bf p})e^{ip_{+}\cdot x}v_{+}^s({\bf p})\right],\nonumber \\
  \psi_ {-} (\vec x,t) &=&\sum_{s}\int \frac{d^3p }{(2\pi)^3}
  \frac{1}{\sqrt{2 E_{-}}} \\
  &&\times \left[b_{-}^{s}({\bf p})e^{-ip_{-}\cdot x}u_{-}^s({\bf p})+
  d^{s\dag}_{-}({\bf p})e^{ip_{-}\cdot x}v_{-}^s({\bf p})\right].\nonumber 
\end{eqnarray}
where $p_{\pm}=(\omega_{\pm},\vec p)$ and $E_{\pm }=\sqrt{\vec{p}^2+m_{ \pm}^2} $ and $u,v$
are the eigenspinors satisfying the orthogonality relations
\begin{eqnarray}
u^{\dag s}_{\pm}u_{\pm}^r=v^{\dag s}_{\pm}v^r_{\pm}=2E_{\pm}\delta^{sr},
\end{eqnarray}
The Hamiltonian of the
theory can be written in terms of the standard creation and
annihilation operators for the fields $\psi_{\pm}$ as
\begin{eqnarray}
H&=&\sum_{s}\int d^3p  \Big(E_{+}(b^{s\dag}_{+}({\vec p})
 b_{+}^s({\vec p})+d^{s\dag}_{+}({\vec p})d^{s}_{+}({\vec p})) \nonumber \\
&& +E_{-}(b^{s\dag}_{-}({\vec p})   b^s_{-}({\vec
p})+d^{s\dag}_{-}({\vec p})d^{s}_{-}({\vec p}))\Big),
\end{eqnarray}
and,
\begin{eqnarray}
\{b^s_{\pm}({\vec p}),b_{\pm}^{r\dag}({\vec k})\}&=&
\pm  (2\pi)^3\delta^{sr}\delta^3({\vec p-\vec k}),\nonumber\\
\{d^s_{\pm}({\vec p}),d_{\pm}^{r\dag}({\vec k})\}&=&\pm (2\pi)^3
\delta^{sr}\delta^3({\vec p - \vec k}),
\end{eqnarray}
are the nonvanishing anticommutators of creation and annihilation
operators for particles (b) and antiparticles (d) of spin $s$ and
$r$. Here the positivity of the energy spectrum and the
indefiniteness of Fock space are evident. The propagators are
\begin{eqnarray}
S_{\pm }(p)= \frac{ \pm i (\sla{p}\pm m_{\pm})}{p^2 - m^2_{\pm }}.
\end{eqnarray}
By introducing interactions the wrong sign may cause the loss of
unitarity. However, it has been shown that with a suitable
prescription for the propagators it is possible to maintain
unitarity \cite{CUTW}. Although unitarity is kept,
causality is lost at a microscopic scale, as can be seen by the
occurrence of negative decay rates.

Summarizing, theories with higher time derivatives have the
following important features (see also \cite{Issues}).
\begin{itemize}
  \item The theory doubles the number of modes.
  \item The new modes correspond to negative norm states.
  \item The theory can always be defined with positive energies
and unitary S matrix.
  \item Causality is lost at a microscopic scale.
\end{itemize}
\section{Quantization}\label{Sec4}
In this section we will proceed to quantize the free Myers-Pospelov
theory for the special case of $n$ purely timelike and $g_2=0$. As
we mentioned above this case corresponds to a higher time derivative
theory and it will have many features in common with the model
reviewed in the previous section. However, we will take a different
strategy for quantizing the theory because our present theory lacks
Lorentz covariance.

In this case the Lagrangian is 
\begin{eqnarray}\label{M-P-Lag}
L=\int d^3x \,\bar \psi(i {\sla{\partial}} - \, g \gamma^0
 \, \partial_t^2-m )\psi,\nonumber \\
= \int d^3x \,\psi^{\dag} (i \partial_t - \, g
\, \partial_t^2-\hat h_D )\psi,
\end{eqnarray}
where $\hat h_D=-i\vec \alpha \cdot \vec \nabla +m\beta$
is the standard Dirac Hamiltonian operator and we have considered 
without loss of generality $g = -g_1$ to make the contact with the previous section more transparent.
Now
let us write the field in terms of the standard solutions
of the Dirac Hamiltonian operator,
\begin{eqnarray}
 \psi({\bf x},t)=\sum_{s,i}\int \frac{d^3p}{(2\pi)^3}\frac{1}{\sqrt {2E_{\vec{p}}}}
 u_i^s(\vec{p}) \psi_i^s(\vec{p},t)e^{i \epsilon_i\vec {p} \cdot \vec x},
\end{eqnarray}
where $s$ is a spin index, $i$ is the particle and antiparticle
index, i.e., $u_1^s=u^s$ and $u_2^s=v^s$ being $u^s$ and $v^s$ the
standard spinors and $E_{\vec{p}}=\sqrt{\vec p^2+m^2}$. Remembering 
\begin{eqnarray}
\hat h_D u^s_i(\vec{p}) e^{i\epsilon_i \vec p \cdot \vec x}=\epsilon_i 
E_{\vec{p}} u^s_i( \vec{p}) e^{i\epsilon_i \vec
p \cdot \vec x},
\end{eqnarray}
with the normalization convention 
\begin{eqnarray}
u^{s \dag}_i (\vec{p})u_j^r(\vec{p})=2E_{\vec{p}}\delta^{sr}\delta_{ij},
\end{eqnarray}
where $\epsilon_1=+1$ and $\epsilon_2=-1$, we have
\begin{eqnarray}
 L=\sum_{s,i}\int d^3p \,\psi_i^{s\dag}(\vec{p},t)(-g
\partial_t^2+i\partial_t-\epsilon_iE_{\vec{p}})\psi^s_i(\vec{p},t).
\end{eqnarray}
Now it is clear we have reduced the quantum field
theory problem to a set of four
quantum mechanical systems at a given momentum.

These quantum mechanical systems have higher time derivatives and
their quantization can be realized in a similar way as it was done
in the previous section, but for a $0+1$ quantum field theory. In
other words for each index $i$ and $s$ we can define the following
fields:
\begin{eqnarray}
\psi^s_{\pm,i}(\vec{p},t)=\beta_i \left(i\partial_t \pm \omega_{ \mp}^{(i)}(\vec{p})\right)
\psi^s_{i}(\vec{p},t),
\end{eqnarray}
where $\beta_{i}=\left(
 \frac{g}{\omega_{+}^{(i)}+\omega_{-}^{(i)}}\right)^{\frac{1}{2}}$ and
\begin{eqnarray} \label{field-energy}
\omega_{\pm}^{(i)}=\frac{\mp 1+\sqrt{1+4g\epsilon_i E_p}}{2g}.
\end{eqnarray}
The Lagrangian in terms of these fields is
\begin{eqnarray} \label{Lmp}
 L&=&\sum_{s,i}\int d^3p \,\psi_{+,i}^{s\dag}
 (i\partial_t- \omega^{(i)}_{+}(p))\psi^{s}_{+,i}\nonumber \\
&&-\sum_{s,i}\int d^3p \,\psi_{-,i}^{s\dag}
(i\partial_t+\omega^{(i)}_{-}(p))\psi^s_{-,i},
\end{eqnarray}
the equations of motion in terms of these fields are
\begin{eqnarray}
(i\partial_t \mp    \omega_{\pm }^{(i)} ) \psi^{s}_{\pm, i}=0,
\end{eqnarray}
whose solution are
\begin{eqnarray}
 \psi^{s}_{ \pm, i}= C^{s}_{ \pm, i}(\vec p) e^{\mp i \omega_{\pm }^{(i)}t  },
\end{eqnarray}
\begin{eqnarray}
 \psi^{\dag  s}_{ \pm, i}= C^{\dag s }_{ \pm, i} (\vec p)e^{\pm i \omega_{\pm }^{(i)}t  }.
\end{eqnarray}
Now it straightforward to quantize this system by promoting the
coefficients $C$ and $C ^\dag $ to operators and taking into account
the minus sign of the second part of (\ref{Lmp}) which produces the
minus sign in the anticommutation relations of the minus modes, i.e,
\begin{eqnarray}\label{antirel}
\{    C^{s}_{ \pm, i}(\vec p), C^{\dag r}_{ \pm, j}(\vec q)\}=\pm (2\pi)^3
\delta^{rs} \delta^3(\vec p-\vec q).
\end{eqnarray}
To make contact with the standard theory note that $C^{s}_{ +, 1} $
corresponds to $b^{s}$ which destroys standard fermion and
 $C^{s}_{ +,2}$ corresponds to $d^{\dag s}$ which creates standard
antifermions, i.e.,
\begin{eqnarray}
 C^{s}_{+,1}({\vec p})  \equiv  b^{s}({\vec p}),
\end{eqnarray}
and
\begin{eqnarray}
 C^{s}_{+,2}({\vec p})   \equiv  d^{\dag s}({\vec p}).
\end{eqnarray}
This correspondence is valid only for the plus modes because the
standard theory is recovered when $g$ goes to zero. However, the
minus modes have not a defining limit and we cannot refer to them
as particle and antiparticle pairs.

The original field can be written as 
\begin{eqnarray}\label{origin-fields}
  \psi(\vec x,t)&=& \psi_ {+} (\vec x,t) -
  \psi_ {-} (\vec x,t), 
\end{eqnarray}
with
\begin{eqnarray}
  \psi_ {+}  (\vec x,t)&=&\sum_{s}\int \frac{d^3p }{(2\pi)^3}\frac{1}
  {\sqrt{2E_{\vec p}}} \\
  &&\times e^{i \vec p \cdot x} \left[\frac{b^s({\bf p})
   u^s({\bf p}) e^{-i\omega_{+}^{(1)}\cdot x}}{(1+4gE_{\vec p})^{1/4}}+
  \frac{
  d^{s\dag}(-{\bf p})v^s({-\bf p})e^{-i \omega_{+}^{(2)}\cdot x}}
  {(1-4gE_{\vec p})^{1/4}}  \right],\nonumber \\
  \psi_ {-} (\vec x,t) &=&\sum_{s}\int \frac{d^3p }{(2\pi)^3}
  \frac{1}{\sqrt{2E_{\vec p}}} \\
  &&\times  e^{i \vec p \cdot x} \left[\frac{C_{-,1}^{s}({\bf p})
  u^s({\bf p})e^{i\omega_{-}^{(1)}\cdot x}}{(1+4gE_{\vec p})^{1/4}}+
  \frac{C^{s\dag}_{-,2}(-{\bf p})v^s(-{\bf p})e^{i \omega_{-}^{(2)}
  \cdot x}}{(1-4gE_{\vec p})^{1/4}}\right].\nonumber 
\end{eqnarray}
Putting it all together from Eq. (\ref{Lmp}) it is easy to arrive
at the expression for the Hamiltonian
\begin{eqnarray}
H&=&\sum_{s}\int d^3p  \Big((\omega_{+}^{(1)} b^{s\dag}({\vec p})
 b^s({\vec p})-\omega_{+}^{(2)} d^{s\dag}({\vec p})d^{s}({\vec p})) \nonumber \\
&& -( \omega_{-}^{(1)}C^{s\dag}_{-,1}({\vec p})   C^s_{-,1}({\vec p})+
\omega_{-}^{(2)} C^{s\dag}_{-,2}({\vec p})C^{s}_{-,2}({\vec p}))\Big).
\end{eqnarray}
The first line of this expression is the standard Hamiltonian in the
limit $g$ goes to zero because $\omega_{+}^{(1)}=-\omega_{+}^{(2)}=E_{\vec p}$. This
Hamiltonian is actually positive if we define the vacuum as the
state which is annihilated by $b$, $d$, $C_{-,1}$ and $C_{-,2}$.
However, in the second line we must use the negativity of the
anticommutators of $C_{-}$ and the positivity of the $\omega_{-}$ to
check this statement.

Now it is clear that the spectrum of the theory is the following:
fermions of spin one half and energy
\begin{eqnarray}
E_{f}=\omega_{+}^{(1)}(\vec p)\approx E_{\vec p}-gE^2_{\vec p},
\end{eqnarray}
and antifermions of spin one half and energy
\begin{eqnarray}
E_{\bar f}=-\omega_{+}^{(2)}(\vec p)\approx E_{\vec p}+gE^2_{\vec p},
\end{eqnarray}
and negative norm particles of spin one half and energies
\begin{eqnarray}
E_{ c}= \omega_{-}^{(1)}(\vec p)\approx \frac{1}{g} +E_{\vec p}-
   gE^2_{\vec p},
\end{eqnarray}
and
\begin{eqnarray}
E_{\bar c}= \omega_{-}^{(2)}(\vec p)\approx \frac{1}{g} -E_{\vec p}-
   gE^2_{\vec p},
\end{eqnarray}
respectively. They sum up for particles of spin one half i.e, eight
modes. This analysis agrees with the discussion in the subsection
(\ref{2a}) restoring $g\to -g_1$ and making the identification 
$\omega_{+}^{(1)}\to \omega_2$, $\omega_{+}^{(2)}\to \omega_1$, 
$\omega_{-}^{(1)}\to-\omega_3$ and $\omega_{-}^{(2)}\to-\omega_4$.
\section{Microcausality}\label{Microcausality}
In this section we will study the source of microcausality violation
due to the noncovariant terms in the model. For this let us 
compute the
anticommutator of free fermionic fields
\begin{eqnarray} 
iS(x-x')=\{\psi(x), \bar \psi(x')\}.
\end{eqnarray}
It is clear from Eqs. (\ref{antirel}) and (\ref{origin-fields}) 
that the plus and minus fields do not mix. Hence,
with $x'=0$ we have
\begin{eqnarray} 
iS(x)&=&\{\psi_{+}(x), \bar \psi_{+}(0)\}+\{\psi_{-}(x), \bar \psi_{-}(0)\}.
\end{eqnarray}
Again restoring $g\to -g_1$ 
the anticommutators can be shown to be
\begin{eqnarray} 
\{\psi_{\pm}(x), \bar \psi_{\pm}(0)\}=(i\sla{\partial}+m) i\Delta_{\pm},
\end{eqnarray}
with
\begin{eqnarray}
\Delta_{+}(x)&=&\int \frac{d^3p}{(2\pi)^32E_{\vec p}} 
e^{i \vec p  \cdot \vec x }
( \frac{e^{-i\omega_1 t}}{\sqrt{1+4g_1E_{\vec p}}}-\nonumber 
\\&& \frac{e^{-i\omega _2 t}}{ \sqrt{1-4g_1E_{\vec p}}}),
\end{eqnarray}
and
\begin{eqnarray}
\Delta_{-}(x)&=&\int \frac{d^3p}{(2\pi)^32E_{\vec p}} 
e^{i\vec p  \cdot \vec x}
( \frac{e^{-i\omega_3 t}}{\sqrt{1-4g_1E_{\vec p}}}-\nonumber
 \\&& \frac{e^{-i\omega _4 t}}{ \sqrt{1+4g_1E_{\vec p}}}),
\end{eqnarray}
where we have used the usual spin sum $\sum_{s}
 u^s(\vec p)\bar u^s(\vec p)=\gamma \cdot p+m $
and $\sum_{s} v^s(\vec p)\bar v^s(\vec p)=\gamma \cdot p-m $. 
Let us combine terms with the same denominator, 
consider thus
\begin{eqnarray}
i\Delta(x)&=&i\Delta_1(x)-i\Delta_2(x),
\end{eqnarray}
where 
\begin{eqnarray}\label{cuteq1}
i\Delta_1(x)&=& \frac{-ie^{-it/2g_1}}{2\pi^2r}  
\int_{0} ^{
\left|\vec{p}\right|_{max}} d \left|{\vec p }\right|\,
 \left|{\vec p }\right| 
\sin(\left|{\vec p }\right|\,r)  \\&&\times \Big(
\frac{e^{\frac{i\sqrt{1-4g_1E_{\vec p}}}{2g_1}t}} {
2E_{\vec p}\sqrt{1-4g_1E_{\vec p}}}-\frac{e^{\frac{-i
\sqrt{1-4g_1E_{\vec p}}}{2g_1}t }} {
2E_{\vec p}\sqrt{1-4g_1E_{\vec p}}} \Big)\nonumber ,
\end{eqnarray}
and
\begin{eqnarray}\label{cuteq}
i\Delta_2(x)&=& \frac{-ie^{-it/2g_1}}{2\pi^2r} \int_{0}^{\infty}
d\left|{\vec p }\right|\, \left|{\vec p }\right| 
\sin(\left|{\vec p }\right|\,r)  \\
&&\times \Big(\frac{e^{\frac
{-i\sqrt{1+4g_1E_{\vec p}} }{2g_1 }t}}{ 2E_{\vec p}
\sqrt{1+4g_1E_{\vec p}}}
 -\frac{e^{\frac{i\sqrt{1+4g_1E_{\vec p}} }{2g_1}t}}
{ 2E_{\vec p}\sqrt{1+4g_1E_{\vec p}}}
\Big),\nonumber
\end{eqnarray}
where $r=\left|{\vec x }\right|$ and we have 
performed the angular integration.

To proceed further let us make the change of
variables $d\left| {\vec p}\right| \left| {\vec p}
\right|=dE_{\vec p} E_{\vec p}$ followed by $z=g_1E_{\vec p}$
to arrive at
\begin{eqnarray}\label{result}
i\Delta_1(x)&=& \frac{e^{-it/2g_1}}{2\pi^2rg_1}
\int_{\epsilon}^{1/4}
dz\, \frac{ \sin (\frac{r \sqrt{z^2-\epsilon^2}}{g_1})
 \sin(\frac{t\sqrt{1-4z} }{2g_1})}
{ \sqrt{1-4z}},
\end{eqnarray}
and
\begin{eqnarray}\label{result1}
i\Delta_2(x)&=& \frac{e^{-it/2g_1}}{2\pi^2rg_1}
 \int_{\epsilon}^{\infty}
dz\,  \frac{\sin(\frac{r\sqrt{z^2-\epsilon^2}}{g_1} )
\sin(\frac{t\sqrt{1+4z}}{2g_1})}{ \sqrt{1+4z}}.
\end{eqnarray}
where $\epsilon=mg_1$.

Alternatively, we could have started with the four momentum
integral representation
\begin{eqnarray}
i\Delta(x)=   \oint_C \frac{d^4p}{(2\pi)^4}
\frac{e^{-ip\cdot x}}
{g^2_1(p_0-\omega_1)(p_0-\omega_2)(p_0-\omega_3)
(p_0-\omega_4)},\nonumber \\
\end{eqnarray}
where $C$ is the contour encircling all the poles in the clockwise
direction and which satisfies
\begin{eqnarray}\label{P-W-J}
iS(x)=  (i\gamma^{\mu}\partial_{\mu}+g_1\gamma^0\,
 \partial_t^2+m) i\Delta(x),
\end{eqnarray}
arriving at the same result as in Eqs. (\ref{result}), (\ref{result1}). 
One advantage, however, is that in this way it is more clear to see that for momenta
higher than $\left|\vec{p}\right|_{max}$ both poles $\omega_2$ and
$\omega_3$ move out from the region enclosed by the contour $C$ and
eventually become purely imaginary, see Fig.\ 3. Hence,
they do not contribute to the integral when $\left|\vec{p}\right|>
\left|\vec{p}\right|_{max}$ producing a natural cutoff in the
integral (\ref{cuteq1}).
\begin{figure}
\centering
\includegraphics[width=0.47\textwidth]{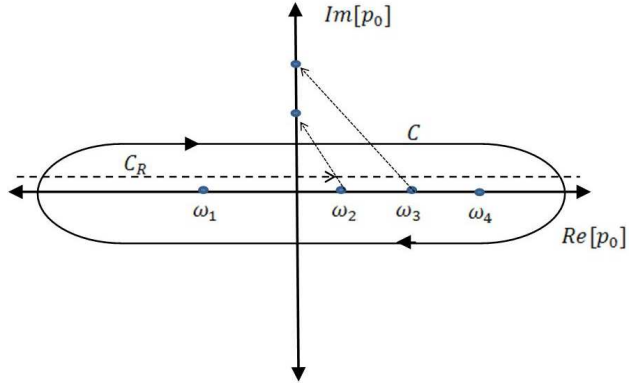}
\caption{For momenta above $\left|\vec{p}\right|_{max}$
both poles $\omega_2$ and $\omega_3$
move out from the region enclosed by
$C$ to the imaginary axis. The dotted contour corresponds to the usual
prescription $C_R$ to be closed from above when $t<0$.}
\end{figure}\label{fig3}

To the lowest order in $\epsilon$ it is possible to solve the
integrals; these are
\begin{eqnarray}
i\Delta_1(x)&=& -\frac{e^{-it/2g_1}}{4(\pi r)^{3/2}\sqrt{2g_1}}  \\
&&\times \left( \cos(\frac{t^2+r^2}{4g_1r})N_1(x,g_1) +
 \sin(\frac{t^2+r^2}{4g_1r})N_2(x,g_1)  \right),\nonumber
\end{eqnarray}
where
\begin{eqnarray}
N_1(x,g_1)={C}(\frac{\alpha r-t}{\sqrt{2 \pi g_1 r}})+2{C}(\frac
{t}{\sqrt{2 \pi g_1 r}})-{C}(\frac{\alpha r+t}
{\sqrt{2 \pi g_1 r}}),\nonumber \\
\end{eqnarray}
\begin{eqnarray}
N_2(x,g_1)={S}(\frac{\alpha r-t}{\sqrt{2 \pi g_1 r}})+2{S}(\frac
{t}{\sqrt{2 \pi g_1 r}})-{S}(\frac{\alpha r+t}
{\sqrt{2 \pi g_1r}}).\nonumber \\
\end{eqnarray}
Above we have introduced the Fresnel integrals  
\begin{eqnarray}
{C}(y)=\int_0^y \cos(\frac{\pi z^2}{2})\, dz,
\end{eqnarray}
\begin{eqnarray}
{S}(y)=\int_0^y \sin(\frac{\pi z^2}{2}) \,dz,
\end{eqnarray}
and defined $\alpha = \sqrt{1-4mg_1}$ and $\beta
= \sqrt{1+4mg_1}$. Similarly, the other part is
\begin{eqnarray}
i\Delta_2(x)&=& -\frac{e^{-it/2g_1}}{4(\pi r)^{3/2}\sqrt{2g_1}} \\
&&\times
  \left( \cos(\frac{t^2+r^2}{4g_1r})N_3(x,g_1) +
 \sin(\frac{t^2+r^2}{4g_1r})N_4(x,g_1)  \right),\nonumber
\end{eqnarray}
with
\begin{eqnarray}
N_3(x,g_1)&=&{C}(\frac{\beta r-t}{\sqrt{2 \pi g_1 r}})-{C}(\frac
{\beta r+t}{\sqrt{2 \pi g_1 r}}),\nonumber
\\
N_4(x,g_1)&=&{S}(\frac{\beta r-t}{\sqrt{2 \pi g_1 r}})-{S}(
\frac{\beta r+t}{\sqrt{2 \pi g_1 r}}).
\end{eqnarray}
For spacelike separations $r^2>t^2$ and making the approximation
for small $g_1$ in order to have $\alpha=\beta \approx 1$ we find
\begin{eqnarray}
i\Delta_1(x)&\to& -\frac{e^{-it/2g_1}}{4(\pi r)^{3/2}
\sqrt{2g_1}}\epsilon(t)  \\
&&\times
\left( \cos(\frac{t^2+r^2}{4g_1r}) +
 \sin(\frac{t^2+r^2}{4g_1r})  \right),\nonumber
\\
i\Delta_2(x)&\to& 0.
\end{eqnarray}
Adding the contributions we have
\begin{eqnarray}
i\Delta(x)&=&-\frac{\epsilon(t)}{8 (\pi r)^{3/2}\sqrt{2g_1}} \\
&& \times
\left(e^{\frac{i(r-t)^2}{4g_1r}}(1-i)+e^{\frac{-i(r+t)^2}{4g_1r}}(1+i)\right),\nonumber
\end{eqnarray}
where $\epsilon(t)=\pm1$ for the corresponding positive and negative values of $t$.
The spacelike regions where microcausality
is violated are the regions where the phase changes slowly:
\begin{eqnarray}
\frac{(r-t)^2}{4g_1r}<0.
\end{eqnarray}
This is
very similar to what occurs in the photon sector of the
Myers-Pospelov theory where the small violations
of microcausality occur near the light cone \cite{11,MR}.
\section{Discussions and conclusions}
In this work we have analyzed some aspects of the fermionic Myers
and Pospelov model: Firstly we have found the general dispersion
relations and solutions of the equation of motion. Secondly we have
analyzed the consistency conditions for the cases purely timelike,
lightlike and purely spacelike. Thirdly we explicitly quantized the
pure time theory and finally we computed the microcausality violation.

In the purely spacelike case no inconsistencies were found. However,
for the other two cases the theory is consistent for momenta below a
natural cutoff. Furthermore, these cases show higher time
derivatives features which double the number of degrees of freedom.
The additional modes are negative norm states which might be
controlled by suitable prescriptions studied in the known Lee-Wick
theories. Microcausality was computed explicitly in the pure time
case, leading to suppressed violations near lightlike four momenta.

In the quantization of the negative norm states appearing in the
theory we have assumed that the Cutkosky prescription should work
for the theory under consideration. However, this is quite far from
being clear, because that procedure was introduced to maintain
unitarity and covariance of Lee-Wick theories. We are not restricted
to fulfill the covariance of the theory but we need to keep
unitarity. This aspect should be studied in future works to complete
the analysis. After this, we would be ready to study new features
due to interaction terms like radiative corrections, the low energy
limit of the theory, and the violation of causality owed to
negative norm states contained in the theory.

The success of the complete answer to these questions would give us
a criteria to establish the validity of the Myers-Pospelov theory as
a consistent effective theory containing possible effects of quantum
gravity.
\begin{acknowledgements}
J. L. acknowledges support from DICYT Grant No. 041131LS (USACH)
and FONDECYT-Chile Grant No. 1100777. C. M. R. acknowledges
partial support from DICYT (USACH) and Direcci\'on de Investigaci\'on de
la Universidad del B\'{\i}o-B\'{\i}o (DIUBB) Grant No. 123809 3/R.
\end{acknowledgements}
\appendix
\section{General solutions and dispersion relations}\label{app}
In this appendix we will characterize the general solutions and
dispersion relations of the equation of motion for the general
fermionic Myers and Pospelov theory. This characterization is not
essential for the understanding of the body of the work apart from
some particular aspects concerning the dispersion relation. However,
we include it for the sake of completeness.

Consider the equation of motion
\begin{eqnarray}
\left(\slashed a-\slashed b\gamma^5 -m \right)\psi=0,
\end{eqnarray}
where $a_\mu \equiv p_\mu-g_1 n_\mu (n\cdot p)^2$ and
$b_\mu\equiv g_2n_\mu (n\cdot p)^2$ are four-vectors which will
help us to clear up the notation.

Let us define the following matrices:
\begin{eqnarray}
\hat { M}\equiv \slashed a-\slashed b\gamma^5 - m, \qquad
\hat h\equiv \left[\slashed a,\slashed b\right]\gamma^5.
\end{eqnarray}
do not confuse the $\hat h$ operator here with the Dirac Hamiltonian
$\hat h_{D}$ in the text. These operators satisfy the relations,
\begin{eqnarray}
[\, \hat{ M},\hat {h}\,]&=&0,\nonumber \\
({\hat {M}}+2m)\,{\hat {M}} &=& a^2 - b^2 - m^2 - \hat h.
\end{eqnarray}
This means that the solutions of the equation of motion can be expressed in terms
of the eigenvectors of $\hat h$.

By noticing that
\begin{eqnarray}
\hat h^2 = 4 \left[ (a\cdot b)^2 - a^2b^2\right],
\end{eqnarray}
the general dispersion relation is given by
\begin{eqnarray} \label{rel-disp3}
\left(a^2-b^2-m^2\right)^2-4\left((a\cdot b)^2-a^2b^2\right)=0,
\end{eqnarray}
or by the Eq. (\ref{reldisp-cuad2}) in terms of $p$.
In the case $b^{\mu}=0$, we have the simplified dispersion relation
\begin{eqnarray}
a^2-m^2 =0.
\end{eqnarray}

Now, we will calculate the solutions of the equations of motion.
As we pointed out above, we can find these solutions among the
eigenvectors $\psi_i$ satisfying,
\begin{eqnarray}
\hat h\psi_i=h_i\psi_i,
\end{eqnarray}
for the eigenvalues $h_i$.
Then, let us find those eigenvectors. To do so,
we notice that the $\hat h$ operator can be written in terms of a rank
two antisymmetric tensor, $T_{\mu\nu}\equiv a_\mu b_\nu-a_\nu b_\mu$, that is,
\begin{eqnarray}
\hat h \equiv T_{\mu\nu}
\epsilon^{\mu\nu\sigma\rho}{\cal S}_{\sigma\rho}.
\end{eqnarray}
with ${\cal S}_{\mu\nu}=\frac{i}{4}\left
[\gamma_\mu,\gamma_\nu\right]$ and the convention $\epsilon^{0123}=1$

From this tensor, we define two orthogonal three-vectors,
\begin{eqnarray}
(\vec u)^i &\equiv&T^{0i}= a^0(\vec b)^i - b^0(\vec a)^i \equiv u \hat e^i_1,  \\
(\vec v)^i &\equiv&\frac{1}2\epsilon^{ijk}T_{jk}= (\vec a\times \vec b)^i \equiv v \hat e^i_2,
\end{eqnarray}
and thus
\begin{eqnarray}
(\vec w)^i &\equiv& (\vec u\times\vec v)^i\equiv u v \hat e^i_3,
\end{eqnarray}
where $\hat e_1$, $\hat e_2 $ and $\hat e_3$
are three orthonormal space vectors on the
direction of $\vec u$, $\vec v$ and $\vec w$,
respectively.
The norm of these vectors are,
\begin{eqnarray}
u &=& \sqrt{(a^0)^2(\vec b)^2 + (b^0)^2(\vec a)^2
- 2(a^0b^0)(\vec a\cdot \vec b)},\nonumber \\
v &=& \sqrt{ (\vec a)^2(\vec b)^2 -(\vec a\cdot\vec b)^2}.
\end{eqnarray}
Note that
\begin{eqnarray}
T^2 &=& T_{\mu\nu}T^{\mu\nu}
= 2(v^2 - u^2), \nonumber \\
&=& 2(a^2 b^2 - (a\cdot b)^2)=-\frac{1}{2} {\hat{h}}^2.
\end{eqnarray}
The negative values of $T^2$ correspond to real
eigenvalues for $\hat h$
and the positive ones correspond to purely imaginary
eigenvalues. By making use
of the analogy with the electromagnetic tensor $F$
we will call the $T^2<0$ 
``electric'' case
and $T^2>0$ the ``magnetic'' case.

Now, we define the rotation and boost generators
in the spinor representation,
\begin{eqnarray}
{\cal J}_i=\frac{1}2 \epsilon_{ijk} {\cal S}^{jk},\quad
{\cal K}_i={\cal S}_{0i},
\end{eqnarray}
where the spatial indices are referring to the $e$
basis defined above.
Then, the $\hat h$ operator turns out to be
\begin{eqnarray}
\hat h = -4 (u{\cal J}_1 + v{\cal K}_2).
\end{eqnarray}
Performing a boost transformation on the eigenspinor
in the $\hat e_3$ direction
\begin{eqnarray}
\psi_h = e^{-i\eta {\cal K}_3}\psi^\prime_h,
\end{eqnarray}
the $\hat h$ operator transforms as
\begin{eqnarray}
\hat h^\prime &\equiv& e^{i\eta{\cal K}_3}\hat h \,e^
{-i\eta{\cal K}_3}=-4\left[(u\cosh\eta-v\sinh\eta){\cal J}_1 \right. \nonumber \\
&& \left.+
(v\cosh\eta - u\sinh\eta){\cal K}_2\right].
\end{eqnarray}
Because $-1<\tanh\eta\,<1$, we can distinguish two cases. For $u>v$
we can set $\tanh\eta = \frac{v}{u}$ so that
\begin{eqnarray}
\hat h^\prime = -4\sqrt{u^2-v^2}{\cal J}_1.
\end{eqnarray}
However, for $v>u$, we can set $\tanh\eta= \frac{u}{v}$
such that
\begin{eqnarray}
\hat h^{\prime} = -4\sqrt{v^2 - u^2}{\cal K}_2.
\end{eqnarray}
Since the eigenvalues of ${\cal J}$ and ${\cal K}$ are $\pm\frac{1}2$
and $\pm\frac{i}2$ respectively, we have
$ h_i=2\epsilon_i \sqrt{u^2-v^2}$ for $u>v$, and
$ h_i=2i\epsilon_i \sqrt{v^2-u^2}$ for $v>u$
as we expected. The convention here is $\epsilon_1=+1$ and $\epsilon_2=-1$.

The eigenspinors in the chiral representation for $u>v$
can be written as
\begin{eqnarray}\label{3}
\psi_i^\prime = \left(\begin{array}{c}\alpha_i \xi_i \\
\beta_i \xi_i\end{array}\right),
\end{eqnarray}
with $(\vec u\cdot\vec\sigma)\xi_i=-\epsilon_i u  \xi_i$.

Notice that these eigenvectors have the property, in the $e$ basis,
\begin{eqnarray}\label{1}
\gamma^1\psi_i= \epsilon_i\gamma^0\gamma^5\psi_i.
\end{eqnarray}
However, for $v>u$, the eigenspinors have the form,
\begin{eqnarray}\label{4}
\psi_i^\prime = \left(\begin{array}{c} \gamma_i\chi_i \\
 \delta_i \sigma^1\chi_i\end{array}\right),
\end{eqnarray}
with $(\vec v\cdot\vec\sigma)\chi_i=\epsilon_i  v  \chi_i$. Similarly,
the eigenspinors have the property, in the $e$ basis,
\begin{eqnarray}\label{2}
\gamma^3\psi_i^\prime= i\epsilon_i\gamma^1\gamma^5
\psi_i^\prime.
\end{eqnarray}
The constants $\alpha_i$, $\beta_i$, $\delta_i$, $\gamma_i$
reflect the fact that the eigenspinors
are twofold degenerate.

Now, we are ready to find the solutions of the equations of motion in terms
of the spinors $\psi_i$.
Performing the same transformation on $\hat{M}$
we obtain after some algebra:

In the electric case ($u>v$), we can set, by choosing appropriately the parameter $\eta$,
$a^\prime_3=b^\prime_3=0$ and find
\begin{eqnarray}
a_0^\prime&=&a_0\sqrt{1-\frac{v^2}{u^2}}=a_0\frac{\vert h\vert}{2u},\nonumber \\
b_0^\prime&=&b_0\sqrt{1-\frac{v^2}{u^2}}=b_0\frac{\vert h\vert}{2u}.
\end{eqnarray}
In the other hand, in the magnetic case ($v>u$),
we can set $a_0^\prime=b_0^\prime=0$ and find
\begin{eqnarray}
a_3^\prime&=&-a_0\sqrt{\frac{v^2}{u^2}-1}=
a_0\frac{\vert h\vert}{2u},\nonumber \\
b_3^\prime &=&-b_0\sqrt{\frac{v^2}{u^2}-1}=
b_0\frac{\vert h \vert }{2u}.
\end{eqnarray}
where $\vert h \vert \equiv \sqrt{ \vert u^2-v^2   \vert  }$
and where we have considered that the
2-direction is perpendicular to $\vec a$
and $\vec b$.
Hence, in the electric case the equations
of motion, $\hat {M}^\prime\psi_i=0$, are
\begin{eqnarray}\label{5}
\left[ \left(a_0^\prime-\epsilon_i b_1\right)
\gamma^0 - \left(b^\prime_0-\epsilon_i
 a_1\right)\gamma^0\gamma^5 - m\right]\psi_i=0,
\end{eqnarray}
where we have used (\ref{1}). This equation
fixes the constants in the Eq. (\ref{3})
\begin{eqnarray}\label{6}
\alpha_i & =& {\cal N}m ,\nonumber \\
\beta_i &=&  {\cal N}\left[(a_0^\prime+
\epsilon_i b_1)-(b_0^\prime+\epsilon_i a_1)\right],
\end{eqnarray}
where ${\cal N}$ is a normalization constant.
The equations of motion, ${M}^\prime\psi_i=0$
in the magnetic case are
\begin{eqnarray}
\left[ (a_1-i\epsilon_i
b_3^\prime)\gamma^1 - (b_1- i\epsilon_i a_3^\prime)\gamma^1
\gamma^5-m\right]\psi_i^\prime=0,
\end{eqnarray}
where we have used property (\ref{2}).
This implies that the constants in the
Eq. (\ref{4}) are
\begin{eqnarray}
\gamma_i & =& {\cal N}^{\prime}m ,
\nonumber \\
\delta_i &=&  {\cal N}^{\prime}\left[(b_1-i
\epsilon_i a^{\prime}_3)-(a_1 -i
\epsilon_i b^{\prime}_3)\right],
\end{eqnarray}
where ${\cal N}^{\prime}$ is another
normalization constant.

\end{document}